\documentclass{mem}
\usepackage{natbib}\usepackage{txfonts}\usepackage{balance}
\usepackage{graphicx}
\usepackage[a4paper,breaklinks,dvipdfm]{hyperref}
\idline{75}{282}
\begin{document}
\def\teff{$T\rm_{eff }$}
\def\kms{$\mathrm {km s}^{-1}$}

\title{
Towards an unbiased stellar census in open clusters using multi-wavelength photometry
}

   \subtitle{}

\author{
F.\,J.\,Galindo-Guil\inst{1}, 
D.\,Barrado\inst{1} 
\and H.\,Bouy\inst{1}
          }

  \offprints{F.\, J.\,  Galindo-Guil}

\institute{
Depto. Astrof\'isica, Centro de Astrobiolog\'ia (INTA-CSIC), 
ESAC campus, 
P.O. Box 78, 28691 Villanueva de la Ca\~nada, Spain\\
\email{pgalindo@cab.inta-csic.es}
}

\authorrunning{Galindo-Guil}

\titlerunning{Towards an unbiased stellar census.}

\abstract{
We look for very low-mass members in open clusters with different ages 
placed at different distances. 
The main goal is to produce a reliable census of low-mass stars for each of the clusters and derive 
the Initial Mass Function. 
To achieve this, we combine deep optical and infrared
photometry from our own observing runs and from different public databases. 
We also characterise the individual stellar parameters of our targets. 

}
\maketitle{}

%%%%%%%%%%%%%%%%%%%%%%%%%%%%%%%%%%%%%%%%%%%%%%%%%%%%%%%%%%%%%%%%%%%%%%%%%%%%%%%%%%%%%%%%%%%%%%%%%%%%%%%%%%%%%%%%%%%%% SECTION introduction 
%%%%%%%%%%%%%%%%%%%%%%%%%%%%%%%%%%%%%%%%%%%%%%%%%%%%%%%%%%%%%%%%%%%%%%%%%%%%%%%%%%%%%%%%%%%%%%%%%%%%%%%%%%%%%%%%%%%%%
%%%%%%%%%%%%%%%%%%%%%%%%%%%%%%%%%%%%%%%%%%%%%%%%%%%%%%%%%%%%%%%%%%%%%%%%%%%%%%%%%%%%%%%%%%%%%%%%%%%%%%%%%%%%%%%%%%%%%
%%%%%%%%%%%%%%%%%%%%%%%%%%%%%%%%%%%%%%%%%%%%%%%%%%%%%%%%%%%%%%%%%%%%%%%%%%%%%%%%%%%%%%%%%%%%%%%%%%%%%%%%%%%%%%%%%%%%%
%%%%%%%%%%%%%%%%%%%%%%%%%%%%%%%%%%%%%%%%%%%%%%%%%%%%%%%%%%%%%%%%%%%%%%%%%%%%%%%%%%%%%%%%%%%%%%%%%%%%%%%%%%%%%%%%%%%%%
%%%%%%%%%%%%%%%%%%%%%%%%%%%%%%%%%%%%%%%%%%%%%%%%%%%%%%%%%%%%%%%%%%%%%%%%%%%%%%%%%%%%%%%%%%%%%%%%%%%%%%%%%%%%%%%%%%%%%
\section{Introduction}
%%%%%%%%%%%%%%%%%%%%%%%%%%%%%%%%%%%%%%%%%%%%%%%%%%%%%%%%%%%%%%%%%%%%%%%%%%%%%%%%%%%%%%%%%%%%%%%%%%%%%%%%%%%%%%%%%%%%% SECTION notes 
%%%%%%%%%%%%%%%%%%%%%%%%%%%%%%%%%%%%%%%%%%%%%%%%%%%%%%%%%%%%%%%%%%%%%%%%%%%%%%%%%%%%%%%%%%%%%%%%%%%%%%%%%%%%%%%%%%%%%
%%%%%%%%%%%%%%%%%%%%%%%%%%%%%%%%%%%%%%%%%%%%%%%%%%%%%%%%%%%%%%%%%%%%%%%%%%%%%%%%%%%%%%%%%%%%%%%%%%%%%%%%%%%%%%%%%%%%%
%%%%%%%%%%%%%%%%%%%%%%%%%%%%%%%%%%%%%%%%%%%%%%%%%%%%%%%%%%%%%%%%%%%%%%%%%%%%%%%%%%%%%%%%%%%%%%%%%%%%%%%%%%%%%%%%%%%%%
%%%%%%%%%%%%%%%%%%%%%%%%%%%%%%%%%%%%%%%%%%%%%%%%%%%%%%%%%%%%%%%%%%%%%%%%%%%%%%%%%%%%%%%%%%%%%%%%%%%%%%%%%%%%%%%%%%%%%
%%%%%%%%%%%%%%%%%%%%%%%%%%%%%%%%%%%%%%%%%%%%%%%%%%%%%%%%%%%%%%%%%%%%%%%%%%%%%%%%%%%%%%%%%%%%%%%%%%%%%%%%%%%%%%%%%%%%%

%    A useful approach to understand the stellar evolution is the study of stellar clusters. 
\textit{A priori}, a stellar association is an homogeneous group of stars, (generally) placed at roughly the same 
distance from us, and formed from the same molecular cloud. 
Our study is focused on young open clusters (OCs). 
Despite the fact that there are several very well studied associations (for instance 
the Pleiades, the Hyades, or M35, see
\citet{stauffer2007}, \citet{perryman1998}, \citet{barrado2001a}, to name a few) 
several questions still remain open:
the properties and evolution of low-mass objects, 
the cluster distances,  
lack of a time scale valid to characterise stellar ages over a wide range and in a homogeneous way,  
or the number of stars in each mass range (best known as Initial Mass Function or IMF for short).
Gaia, the European Space Agency mission, will determine cluster distances and 
very accurate proper motions for a large amount of clusters, and 
will identify a significant number of low-mass cluster members.
Even though additional data are needed to reach the low-mass regime in a large sample 
of OCs and try to answer the previous questions. 

Our goal is to complement Gaia's result by looking 
for very low-mass members in a sample of open clusters, 
with several ages and distances located at different regions of the Galaxy.
We will produce a reliable census for each cluster, focusing on the low-mass 
regime near the substellar frontier. 
In this paper we describe our project and the current status.

% The OCs sample and the photometry set used
% is described in Section 1. 
% Section 2 shows how to assess potential candidates for each OC 
% and the estimation of their stellar parameters. 

%%%%%%%%%%%%%%%%%%%%%%%%%%%%%%%%%%%%%%%%%%%%%%%%%%%%%%%%%%%%%%%%%%%%%%%%%%%%%%%%%%%%%%%%%%%%%%%%%%%%%%%%%%%%%%%%%%%%% SECTION objective_and_sample 
%%%%%%%%%%%%%%%%%%%%%%%%%%%%%%%%%%%%%%%%%%%%%%%%%%%%%%%%%%%%%%%%%%%%%%%%%%%%%%%%%%%%%%%%%%%%%%%%%%%%%%%%%%%%%%%%%%%%%
%%%%%%%%%%%%%%%%%%%%%%%%%%%%%%%%%%%%%%%%%%%%%%%%%%%%%%%%%%%%%%%%%%%%%%%%%%%%%%%%%%%%%%%%%%%%%%%%%%%%%%%%%%%%%%%%%%%%%
%%%%%%%%%%%%%%%%%%%%%%%%%%%%%%%%%%%%%%%%%%%%%%%%%%%%%%%%%%%%%%%%%%%%%%%%%%%%%%%%%%%%%%%%%%%%%%%%%%%%%%%%%%%%%%%%%%%%%
%%%%%%%%%%%%%%%%%%%%%%%%%%%%%%%%%%%%%%%%%%%%%%%%%%%%%%%%%%%%%%%%%%%%%%%%%%%%%%%%%%%%%%%%%%%%%%%%%%%%%%%%%%%%%%%%%%%%%
%%%%%%%%%%%%%%%%%%%%%%%%%%%%%%%%%%%%%%%%%%%%%%%%%%%%%%%%%%%%%%%%%%%%%%%%%%%%%%%%%%%%%%%%%%%%%%%%%%%%%%%%%%%%%%%%%%%%%
\section{Objectives and sample}
%%%%%%%%%%%%%%%%%%%%%%%%%%%%%%%%%%%%%%%%%%%%%%%%%%%%%%%%%%%%%%%%%%%%%%%%%%%%%%%%%%%%%%%%%%%%%%%%%%%%%%%%%%%%%%%%%%%%% SECTION objective_and_sample 
%%%%%%%%%%%%%%%%%%%%%%%%%%%%%%%%%%%%%%%%%%%%%%%%%%%%%%%%%%%%%%%%%%%%%%%%%%%%%%%%%%%%%%%%%%%%%%%%%%%%%%%%%%%%%%%%%%%%%
%%%%%%%%%%%%%%%%%%%%%%%%%%%%%%%%%%%%%%%%%%%%%%%%%%%%%%%%%%%%%%%%%%%%%%%%%%%%%%%%%%%%%%%%%%%%%%%%%%%%%%%%%%%%%%%%%%%%%
%%%%%%%%%%%%%%%%%%%%%%%%%%%%%%%%%%%%%%%%%%%%%%%%%%%%%%%%%%%%%%%%%%%%%%%%%%%%%%%%%%%%%%%%%%%%%%%%%%%%%%%%%%%%%%%%%%%%%
%%%%%%%%%%%%%%%%%%%%%%%%%%%%%%%%%%%%%%%%%%%%%%%%%%%%%%%%%%%%%%%%%%%%%%%%%%%%%%%%%%%%%%%%%%%%%%%%%%%%%%%%%%%%%%%%%%%%%
%%%%%%%%%%%%%%%%%%%%%%%%%%%%%%%%%%%%%%%%%%%%%%%%%%%%%%%%%%%%%%%%%%%%%%%%%%%%%%%%%%%%%%%%%%%%%%%%%%%%%%%%%%%%%%%%%%%%%

   The targets of our study are young open clusters of different ages and environments. 
Photometric and spectroscopic data are used in the analysis, and in some cases astrometry measurements. 
Once the census is completed, we will study the IMFs, 
the structure of the clusters and, if possible, derive ages with different techniques. 

   The first step is set up a suitable sample of OCs ready for study. 
We select those OCs visible from the northern hemisphere, plus NGC2451A and B,
with distances and ages up to 500~pc and 500~Myr, 
respectively. The selection is made it using the \citet{dias2002e} catalogue. 
We retrieved a total of 12 targets with a variety of 
distances and ages ranging 190~pc to 400~pc and 20~Myr to 400~Myr. 
The exception is M36, located at about 1330~pc and an age of 25~Myr (see Table \ref{tab:remarks_clusters}).
% The selected sample is showed in Figure \ref{fig:GalacticCoordinates}.

% % FIGURE 1
% %-------------------------------------------------------------
% \begin{figure*}[t!]
%   \resizebox{\hsize}{!}{\includegraphics[clip=true]{GalacticCoordinates_OClustes.eps}}
%   \caption{\footnotesize
%            Merged images from the Planck/HFI instrument, in the frequencies 
%            353~GHz, 545~GHz and 857~GHz for all sky retrieve 
%            from Aladin \citep{bonnarel2000}. 
%            The clusters are plotted with red filled circles. 
%            The sizes do not correspond to the apparent sizes of the clusters.
% %            The large of the clusters are placed in the Galactic Disc.
%            }
%   \label{fig:GalacticCoordinates}
% \end{figure*}
% %-------------------------------------------------------------  

For each OC, 
we gathered all the available information: members and candidates from other works, 
distances, ages, radial velocities, $E(B-V)$, and metallicities.
The WEBDA\footnote{http://www.univie.ac.at/webda/} database, SIMBAD \citep{egret1991} and \citet{kharchenko2005}, 
have been used for this task.

% TABLE 1
%-------------------------------------------------------------
\begin{table*}
 \begin{center}
 \caption{Basic data for our sample of open clusters.}
 \label{tab:remarks_clusters}
   \scriptsize{
      \begin{tabular}{l|cccccccc}
\hline \hline

CLUSTER    & RA J2000 & Dec J2000 &Distance& $E(B-V)$ $/$   &  $logt$ $/$ & Diameter &Photometry& $Gaia$  Mass \\
           &  (h m s) &  (d m s)  & (pc)   & $A_{V}$ (mag)  &  Age (Myr)  & (arcmin) &          & $M (M_{\odot})$  \\
\hline
NGC1960/M36& 05 36 18 & +34 08 24 & 1330  &  0.22 / 0.704   &   7.4/25    &   10     & INT-WFC, $JKs$-KPNO  & 0.42 \\
           &          &           &       &                 &             &          &           &\\ 
NGC 7058   & 21 21 53 & +50 49 11 &  400  &  0.06 / 0.1932  & 8.35 /223   &    7     & INT-WFC, $IZ$-CFHT12K, SDSS & 0.17 \\ 
           &          &           &       &                 &             &          &           &\\                   
Stock 23   & 03 16 11 & +60 02 59 &  380  &  0.26 / 0.8372  & 7.51 / 32   &   29.0   & INT-WFC   &                   0.11 \\ 
           &          &           &       &                 &             &          &           &\\          
ASCC 127   & 23 08 24 & +64 51 00 &  350  & 0.10 / 0.322    & 7.82 / 66   &   86.4   & INT-WFC, SDSS               & 0.11 \\
           &          &           &       &                 &             &          &           &\\  
Stock 10   & 05 39 00 & +37 56 00 &  380  & 0.07 / 0.2254   & 7.9  / 79   &   25     & INT-WFC   &                   0.085\\
           &          &           &       &                 &             &          &           &\\            
ASCC 123   & 22 42 35 & +54 15 35 &  250  & 0.10 / 0.322    & 8.41 / 257  &  153.6   & INT-WFC, SDSS               & 0.13 \\ 
           &          &           &       &                 &             &          &           &\\            
NGC7092/M39& 21 31 48 & +48 26 00 &  326  & 0.013 / 0.04186 & 8.445/ 279  &   29.0   & INT-WFC, $UBV$-KPNO   & 0.14 \\
           &          &           &       &                 &             &          &           &\\            
Platais 2  & 01 13 50 & +32 01 42 &  201  & 0.05 / 0.161    & 8.6 /  398  &   336    & SDSS      &                   0.11 \\
           &          &           &       &                 &             &          &           &\\            
Herschel 1 & 07 47 02 & +00 01 06 &  370  & 0.02 / 0.644    & 8.44 / 275  &   43.2   & SDSS      &                   0.19 \\
           &          &           &       &                 &             &          &           &\\            
NGC 2451 A & 07 43 12 & -38 24 00 &  189  & 0.01 / 0.0322   & 7.78 / 60   &   120    & $BVI$ $Ic/Iwp$ ESO-WFI      & 0.059\\
           &          &           &       &                 &             &          &           &\\            
NGC 2451 B & 07 44 27 & -37 40 00 &  302  & 0.055 / 0.176   & 7.648 / 44  &   180    & $BVI$ $Ic/Iwp$ ESO-WFI      & 0.081\\
           &          &           &       &                 &             &          &           &\\            
ASCC 20    & 05 28 44 & +01 37 48 &  450  &  0.04 / 0.1284  &  7.35 / 22  &  90.0    & SDSS                        & 0.091\\
\hline\hline
        \end{tabular}
     }
\end{center}   
\textbf{Note}: For all OCs we also have retrieved photometry from the catalogues: 
Tycho-2, %\citep{hog2000},
UCAC 4, %\citep{zacharias2013},
2MASS, %\citep{cutri2003},
and 
WISE. %\citep{cutri2012}.
The last column is an estimation of the mass of the faintest object observed with $Gaia$, 
The mass has been calculated assuming distances, ages and $A_{v}$ of the table for each cluster 
at magnitude $G=20$~mag, -the stellar survey will be complete to magnitude $G=20$~mag, \citep{luisma2013}-  
using BT-Settl models.
\end{table*}
%-------------------------------------------------------------

%%%%%%%%%%%%%%%%%%%%%%%%%%%%%%%%%%%%%%%%%%%%%%%%%%%%%%%%%%%%%%%%%%%%% SUBSECTION phot_data
%%%%%%%%%%%%%%%%%%%%%%%%%%%%%%%%%%%%%%%%%%%%%%%%%%%%%%%%%%%%%%%%%%%%%
%%%%%%%%%%%%%%%%%%%%%%%%%%%%%%%%%%%%%%%%%%%%%%%%%%%%%%%%%%%%%%%%%%%%%
\subsection{Photometry data}
%%%%%%%%%%%%%%%%%%%%%%%%%%%%%%%%%%%%%%%%%%%%%%%%%%%%%%%%%%%%%%%%%%%%% 
%%%%%%%%%%%%%%%%%%%%%%%%%%%%%%%%%%%%%%%%%%%%%%%%%%%%%%%%%%%%%%%%%%%%%
%%%%%%%%%%%%%%%%%%%%%%%%%%%%%%%%%%%%%%%%%%%%%%%%%%%%%%%%%%%%%%%%%%%%%

The photometric datasets are from optical (for example
SDSS) to mid-infrared and varies for each cluster.
% Photometry ranges from optical (for example SDSS) up to mid-infrared. 
% The photometric set is different for each cluster. 
Some bands are commons for all OCs (2MASS and WISE), 
whereas others are unique for each association. 
% The set is showed in Table\ref{tab:remarks_clusters}. 
Photometric data not coming from public surveys has been reduced 
in the same way as DANCe project \citep{hbouy2013}. 

\begin{figure}[]
  \resizebox{\hsize}{!}{\includegraphics[clip=true]{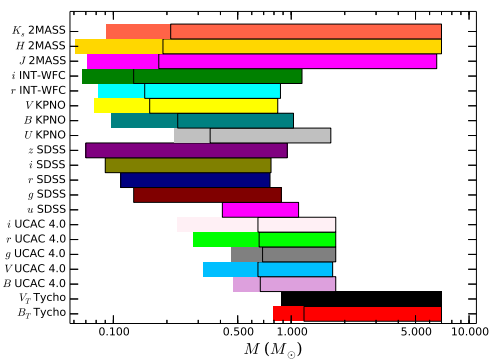}}
  \caption{\footnotesize
           The diagram shows the mass range covered for all 
           photometric bands for the cluster M39. 
           The saturation, completeness and limiting mass for each band 
           have been calculated 
           with the Lyon models \citep{allard2012} or Siess models \citep{siess2000}. 
           The black edge rectangles show the mass range coverages between saturation 
           completeness 
           assuming a distance of 303~pc, $A_{v}=$~0.064~mag an age of 280~Myr \citep{vanleeuwen2009}.
          }
  \label{fig:plot_mass_range}
\end{figure}

Due to the variety of photometry, the lowest mass observed 
in each cluster is different. 
It is important to know how deep is the photometry in each band 
and the area in the sky covered by it, 
so we can construct the appropriate Colour Magnitude Diagrams (CMDs) 
and Colour Colour Diagrams (CCDs) to select candidates. 
An example is shown in Figure \ref{fig:plot_mass_range} that contains 
the mass range covered by each band in M39.
In this case the potential minimum mass for a cluster member is 
roughly 0.061~$M_{\odot}$.
% Saturation, completeness and detection limits have been considered in the calculation 
% assuming a distance of 303~pc, $A_{v}=$~0.064~mag and an age of 280~Myr \citep{vanleeuwen2009}. 

% FIGURE 2
%-------------------------------------------------------------
%-------------------------------------------------------------  

%%%%%%%%%%%%%%%%%%%%%%%%%%%%%%%%%%%%%%%%%%%%%%%%%%%%%%%%%%%%%%%%%%%%%%%%%%%%%%%%%%%%%%%%%%%%%%%%%%%%%%%%%%%%%%%%%%%%% SECTION selection_candidates 
%%%%%%%%%%%%%%%%%%%%%%%%%%%%%%%%%%%%%%%%%%%%%%%%%%%%%%%%%%%%%%%%%%%%%%%%%%%%%%%%%%%%%%%%%%%%%%%%%%%%%%%%%%%%%%%%%%%%%
%%%%%%%%%%%%%%%%%%%%%%%%%%%%%%%%%%%%%%%%%%%%%%%%%%%%%%%%%%%%%%%%%%%%%%%%%%%%%%%%%%%%%%%%%%%%%%%%%%%%%%%%%%%%%%%%%%%%%
%%%%%%%%%%%%%%%%%%%%%%%%%%%%%%%%%%%%%%%%%%%%%%%%%%%%%%%%%%%%%%%%%%%%%%%%%%%%%%%%%%%%%%%%%%%%%%%%%%%%%%%%%%%%%%%%%%%%%
%%%%%%%%%%%%%%%%%%%%%%%%%%%%%%%%%%%%%%%%%%%%%%%%%%%%%%%%%%%%%%%%%%%%%%%%%%%%%%%%%%%%%%%%%%%%%%%%%%%%%%%%%%%%%%%%%%%%%
%%%%%%%%%%%%%%%%%%%%%%%%%%%%%%%%%%%%%%%%%%%%%%%%%%%%%%%%%%%%%%%%%%%%%%%%%%%%%%%%%%%%%%%%%%%%%%%%%%%%%%%%%%%%%%%%%%%%%
\section{Selection of candidate members}
%%%%%%%%%%%%%%%%%%%%%%%%%%%%%%%%%%%%%%%%%%%%%%%%%%%%%%%%%%%%%%%%%%%%%%%%%%%%%%%%%%%%%%%%%%%%%%%%%%%%%%%%%%%%%%%%%%%%% SECTION selection_candidates 
%%%%%%%%%%%%%%%%%%%%%%%%%%%%%%%%%%%%%%%%%%%%%%%%%%%%%%%%%%%%%%%%%%%%%%%%%%%%%%%%%%%%%%%%%%%%%%%%%%%%%%%%%%%%%%%%%%%%%
%%%%%%%%%%%%%%%%%%%%%%%%%%%%%%%%%%%%%%%%%%%%%%%%%%%%%%%%%%%%%%%%%%%%%%%%%%%%%%%%%%%%%%%%%%%%%%%%%%%%%%%%%%%%%%%%%%%%%
%%%%%%%%%%%%%%%%%%%%%%%%%%%%%%%%%%%%%%%%%%%%%%%%%%%%%%%%%%%%%%%%%%%%%%%%%%%%%%%%%%%%%%%%%%%%%%%%%%%%%%%%%%%%%%%%%%%%%
%%%%%%%%%%%%%%%%%%%%%%%%%%%%%%%%%%%%%%%%%%%%%%%%%%%%%%%%%%%%%%%%%%%%%%%%%%%%%%%%%%%%%%%%%%%%%%%%%%%%%%%%%%%%%%%%%%%%%
%%%%%%%%%%%%%%%%%%%%%%%%%%%%%%%%%%%%%%%%%%%%%%%%%%%%%%%%%%%%%%%%%%%%%%%%%%%%%%%%%%%%%%%%%%%%%%%%%%%%%%%%%%%%%%%%%%%%%

   First of all we have to recover previous members (if that is the case) and placed 
them in several CMDs.
This gives us an idea of the sequence in the brighter regime. 
We consider as candidates those sources located above an isochrone, and no candidates those 
placed below it.
We would like to point out that we weigh up the uncertainties in the photometry and 
keeping in mind the saturation, completeness and detection limits of each band. 

The selection is made with the appropriate photometric filters for the BT-Settl \citep{allard2012} models
and the \citet{siess2000} evolutionary tracks. So the selection covers a wide range of masses.  
From all the ages recovered from the literature we choose the oldest one, and shift at the farthest 
distance considering the cluster reddening. The Spanish Virtual Observatory provides a 
Filter Profile Service\footnote{http://svo2.cab.inta-csic.es/theory/fps/index.php?mode=voservice}, 
an useful tool to transform $A_{V}$ into the absorption in an each specific photometric band. 

The final selection has several categories of candidates: 
probable candidates -sources flagged as candidates in all CMDs-, 
probable no candidates -sources flagged as not candidates in at least one CMD-, 
possible candidates -sources flagged as members in the deepest CMDs and not detected in brighter ones-.
% (see Figure \ref{fig:CMD_ascc127}).
% is the final result from the selection process.

% FIGURE 3
%-------------------------------------------------------------
\begin{figure}[]
  \resizebox{\hsize}{!}{\includegraphics[clip=true]{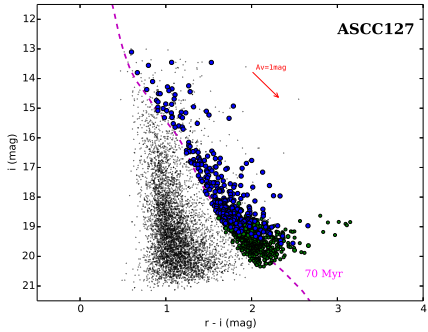}}
  \caption{\footnotesize
           The CMD ($(r-i)$, $i$) reaches the lowest mass in ASCC127, 
           and it shows the final selection. 
           Blue filled circles are probable candidates in all CMDs and CCDs, 
           green points are possible candidates that are not detect in all CMDs/CCDs. 
           The magenta dash-line is a 70~Myr isochrone \citep{allard2012} 
           shifted at the distance and $A_{v}$ of the cluster 
           (see Table \ref{tab:remarks_clusters}).
           }
  \label{fig:CMD_ascc127}
\end{figure}
%-------------------------------------------------------------  

%%%%%%%%%%%%%%%%%%%%%%%%%%%%%%%%%%%%%%%%%%%%%%%%%%%%%%%%%%%%%%%%%%%%% SUBSECTION stellar_parameters
%%%%%%%%%%%%%%%%%%%%%%%%%%%%%%%%%%%%%%%%%%%%%%%%%%%%%%%%%%%%%%%%%%%%%
%%%%%%%%%%%%%%%%%%%%%%%%%%%%%%%%%%%%%%%%%%%%%%%%%%%%%%%%%%%%%%%%%%%%%
\subsection{Stellar Parameters}
%%%%%%%%%%%%%%%%%%%%%%%%%%%%%%%%%%%%%%%%%%%%%%%%%%%%%%%%%%%%%%%%%%%%% 
%%%%%%%%%%%%%%%%%%%%%%%%%%%%%%%%%%%%%%%%%%%%%%%%%%%%%%%%%%%%%%%%%%%%%
%%%%%%%%%%%%%%%%%%%%%%%%%%%%%%%%%%%%%%%%%%%%%%%%%%%%%%%%%%%%%%%%%%%%%
   
   Our next step is to characterise the stellar parameters 
from our potential candidates, in particular 
bolometric luminosities ($L_{bol}$) and effective temperatures ($T_{\rm{eff}}$). 

\begin{figure}[]
  \resizebox{\hsize}{!}{\includegraphics[clip=true]{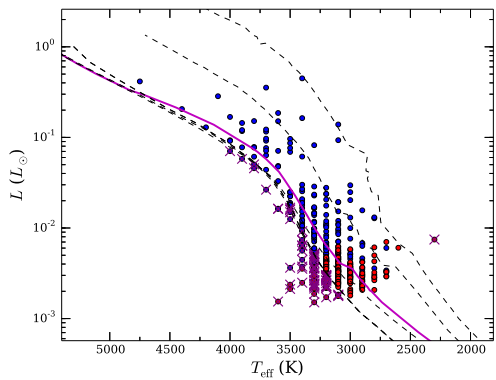}}
  \caption{\footnotesize
           HRD for ASCC127. Solid circles are the candidates: 
           in blue those with photometric measurements in all
           bands, in red candidates with missing photometry.
           Dash-lines correspond to 1-10-100-1000-10000~Myr
           isochrones \citep{allard2012}. The magenta line correspond to 100 Myr.
           The rejected members below the Main Sequence have 
           over-imposed big magenta crosses. 
           }
  \label{fig:HRD_ascc127}
\end{figure}

Deriving just $L_{bol}$ and $T_{\rm{eff}}$ building a 
Spectral Energy Distribution (SED) is more robust than using one colour index.
In this way, 
we eliminate the uncertainties in the luminosities that appear 
when bolometric corrections are used.
Our multi-wavelength approach allows us to build reliable SEDs and
use VOSA \citep{bayo2008}
to obtain $L_{bol}$ and $T_{\rm{eff}}$.
In particular, since in several cases we have a significant number 
of datapoints covering a large range in wavelength, the errors in 
$L_{bol}$ are very reduced.
VOSA can also estimates masses and ages for each member 
using several grids of models (\citet{allard2012}, \citet{siess2000}).
We generate a Hertzsprung-Russell Diagram (HRD) and reject additional possible 
non-candidates with the values of $T_{\rm{eff}}$ and the $L_{bol}$. 
The HRD for ASCC127 with potential candidates is shown 
in Figure \ref{fig:HRD_ascc127}.

% FIGURE 3
%-------------------------------------------------------------
%-------------------------------------------------------------  

%%%%%%%%%%%%%%%%%%%%%%%%%%%%%%%%%%%%%%%%%%%%%%%%%%%%%%%%%%%%%%%%%%%%%%%%%%%%%%%%%%%%%%%%%%%%%%%%%%%%%%%%%%%%%%%%%%%%% SECTION future 
%%%%%%%%%%%%%%%%%%%%%%%%%%%%%%%%%%%%%%%%%%%%%%%%%%%%%%%%%%%%%%%%%%%%%%%%%%%%%%%%%%%%%%%%%%%%%%%%%%%%%%%%%%%%%%%%%%%%%
%%%%%%%%%%%%%%%%%%%%%%%%%%%%%%%%%%%%%%%%%%%%%%%%%%%%%%%%%%%%%%%%%%%%%%%%%%%%%%%%%%%%%%%%%%%%%%%%%%%%%%%%%%%%%%%%%%%%%
%%%%%%%%%%%%%%%%%%%%%%%%%%%%%%%%%%%%%%%%%%%%%%%%%%%%%%%%%%%%%%%%%%%%%%%%%%%%%%%%%%%%%%%%%%%%%%%%%%%%%%%%%%%%%%%%%%%%%
%%%%%%%%%%%%%%%%%%%%%%%%%%%%%%%%%%%%%%%%%%%%%%%%%%%%%%%%%%%%%%%%%%%%%%%%%%%%%%%%%%%%%%%%%%%%%%%%%%%%%%%%%%%%%%%%%%%%%
%%%%%%%%%%%%%%%%%%%%%%%%%%%%%%%%%%%%%%%%%%%%%%%%%%%%%%%%%%%%%%%%%%%%%%%%%%%%%%%%%%%%%%%%%%%%%%%%%%%%%%%%%%%%%%%%%%%%%
\section{Future work}
%%%%%%%%%%%%%%%%%%%%%%%%%%%%%%%%%%%%%%%%%%%%%%%%%%%%%%%%%%%%%%%%%%%%%%%%%%%%%%%%%%%%%%%%%%%%%%%%%%%%%%%%%%%%%%%%%%%%% SECTION future 
%%%%%%%%%%%%%%%%%%%%%%%%%%%%%%%%%%%%%%%%%%%%%%%%%%%%%%%%%%%%%%%%%%%%%%%%%%%%%%%%%%%%%%%%%%%%%%%%%%%%%%%%%%%%%%%%%%%%%
%%%%%%%%%%%%%%%%%%%%%%%%%%%%%%%%%%%%%%%%%%%%%%%%%%%%%%%%%%%%%%%%%%%%%%%%%%%%%%%%%%%%%%%%%%%%%%%%%%%%%%%%%%%%%%%%%%%%%
%%%%%%%%%%%%%%%%%%%%%%%%%%%%%%%%%%%%%%%%%%%%%%%%%%%%%%%%%%%%%%%%%%%%%%%%%%%%%%%%%%%%%%%%%%%%%%%%%%%%%%%%%%%%%%%%%%%%%
%%%%%%%%%%%%%%%%%%%%%%%%%%%%%%%%%%%%%%%%%%%%%%%%%%%%%%%%%%%%%%%%%%%%%%%%%%%%%%%%%%%%%%%%%%%%%%%%%%%%%%%%%%%%%%%%%%%%%
%%%%%%%%%%%%%%%%%%%%%%%%%%%%%%%%%%%%%%%%%%%%%%%%%%%%%%%%%%%%%%%%%%%%%%%%%%%%%%%%%%%%%%%%%%%%%%%%%%%%%%%%%%%%%%%%%%%%%

   The current status of our project has been described. 
The multi-wavelength approach allows us to derive a reliable census 
down to low-mass candidates.
Gaia will provide exact distances and proper motions of these clusters, 
although our survey is deeper and will complement those data. 

In addition, we are implementing statistically robust selection methods 
\citep{luisma2014}, following the same methodology as the DANCe project 
regarding the photometry  -\citet{hbouy2013}, \citet{hbouy2014} and \citet{hbouy2014b}-. 
When it is feasible, we will also use proper motions, mainly to reject foreground 
objects.
The subsequent step would be to derive IMFs and study the spatial distribution in a wide mass range and 
several environments.  
Different observing campaigns have been carried out to confirm candidates with spectroscopy 
and derive spectral types to estimate the loci of the lithium depletion boundary.

\begin{acknowledgements}
FJGG thanks Benjam\'in Montesinos 
and Mar\'ia Morales-Calder\'on for their comments and suggestions; 
the SOC/LOC of this congress and specially Ricky Smart for the support
and for the rewarding atmosphere; 
France Allard provided evolutionary stellar tracks, 
This research has been funded by Spanish grant AYA2010-21161-C02-02
and AYA2012-38897-C02-01. 
% WEBDA
It has made use of the WEBDA database, 
operated at the Department of Theoretical Physics and 
Astrophysics of the Masaryk University.
%% VOSA
%This publication makes use of VOSA, 
%developed under the Spanish Virtual Observatory project 
%supported from the Spanish MICINN through grant AyA2008-02156.

\end{acknowledgements}

\bibliographystyle{aa}
%\bibliography{/pcdisk/einstein/pgalindo/data/010_WRITTING/Bibliografia}

\end{document}